\documentclass[twocolumn, prl, showkeys, showpacs, amsmath,amssymb]{revtex4}

\usepackage{graphicx}
\usepackage{dcolumn}
\usepackage{bm}
\usepackage[cp1250]{inputenc}
\usepackage[usenames,dvipsnames]{color}
\usepackage{ifthen} 
\usepackage{amssymb,amstext,amsfonts,amsmath}
\begin{document}

\newcommand{\tr}{\textcolor{Red}}

\newcommand{\mb}{\mathbf}

\preprint{APS/123-QED}



\title{Exact solutions of kinetic equations in an autocatalytic growth model}

\author{Jakub J\c{e}drak} 
\email[electronic address: ]{jedrak@agh.edu.pl} 
\affiliation{Faculty of Non-Ferrous Metals, AGH University of Science and Technology,  Al. Mickiewicza 30, 30-059 Cracow, Poland}

\date{\today}

\begin{abstract} 
Kinetic equations are introduced for the transition-metal nanocluster nucleation and growth mechanism, as proposed by Watzky and Finke. Equations of this type take the form of Smoluchowski coagulation equations supplemented with the terms responsible for the chemical reactions. In the absence of coagulation, we find complete analytical solutions of the model equations for the autocatalytic rate constant both proportional to the cluster mass, and the mass-independent one. In the former case, $\xi_{k} = s_k(\xi_{1})\propto \xi_{1}^{k}/k $ was obtained, while in the latter, the functional form of $s_k(\xi_{1})$ is more complicated. In both cases, $\xi_{1}(t) = h_{\mu}(M_{\mu}(t))$ is a function of the moments of the mass distribution. 
Both functions, $s_k(\xi_{1})$ and $h_{\mu}(M_{\mu})$, depend on the assumed mechanism of autocatalytic growth and monomer production, and not on other chemical reactions present in a system.

\end{abstract} 
\maketitle

Nucleation and growth phenomena, resulting in occurrence of a new phase from a homogeneous host phase, are ubiquitous in nature \cite{Topics in, SDH}. In many cases, apart from processes of coagulation and fragmentation, chemical reactions are also present in a system. Frequently, the chemical reactions account for phenomena studied by polymer and colloidal science.  

Nucleation and subsequent growth of metal nanoclusters in aqueous solution is a topic of considerable current interest, since solution route synthesis is one of the most convenient methods of producing transition-metal nanoparticles \cite{nano}. However,  applicability of this method depends on the ability to control size and shape of the produced nanoparticles, which determine their unique optical, electronic and catalytic properties. For that reason, theoretical models capable of predicting the cluster size distribution, as well as its dependence on the experimentally controllable parameters of the system, are required.

A mechanism of transition-metal colloidal nanoparticle formation has been proposed by Watzky and Finke \cite{WF 1}, cf. \cite{WF 2, WF 3, WF 4, WF protein aggregation}. The WF mechanism  consists of (i) slow monomer, i.e., the zerovalent transition-metal atom ($\text{B}_1$) production due to reduction reaction $\text{A} \rightarrow \text{B}_1$ of a metal precursor ($\text{A}$), usually a transition-metal complex coordination compound, (ii) fast autocatalytic reduction reaction $\text{A} + \text{B}_i  \rightarrow \text{B}_{i+1}$ taking place on the surface of growing metal nanoparticles, consisting of $i$ atoms ($\text{B}_i$), and (iii)  process of coagulation $\text{B}_i + \text{B}_j \rightleftarrows \text{B}_{i+j}$, reversible or otherwise. In the original WF scheme \cite{WF 1}, step (iii) had not been considered. Irreversible coagulation was first introduced in \cite{WF 2}.

For the transition metals in which higher oxidation states are present, e.g. $\text{Au}$, at least one additional preliminary step of the form (iv) $\text{P} \rightarrow \text{A}$ is needed \cite{Paclawski Fitzner gold, Tatarchuk et al, Our group ascorbic acid, BWKEIK, Our group glucose}. 

As an excess of the reducing agent is usually used in the reactions (i), (ii), (iv), its concentration is fairly time-independent, and all chemical reactions may be treated as irreversible. Furthermore, (i) and (iv) may be treated as reactions of pseudo-first order, while (ii) as a reaction of pseudo-second order.

The WF mechanism is applicable to other systems. Certain cases of transition metal oxides or sulfides nanocluster formation, or polymerization phenomena of various kind, including protein aggregation \cite{WF protein aggregation}, are well described by an effective model defined by (i)-(iv), even if the actual mechanism of nucleation and growth in these systems is more complicated.

However, the kinetic equations corresponding to step (iii), proposed in \cite{WF 2, WF 3, WF 4}, cannot be regarded as fully satisfactory. In particular, the original approach does not allow to predict the cluster size distribution. Thus, in this Rapid Communication proper kinetic rate equations for the WF mechanism are introduced.

\textit{Kinetic rate equations.} In order to describe kinetics of step (iii), the approach proposed by Smoluchowski \cite{Smoluchowski original, Aldous}, a standard and widely-used tool for description of various coagulation phenomena, is reintroduced \cite{Hendriks Ernst, Leyvraz solo, PRL 1, PRL 3, PRB R 1, PRA 1, J PHYS A 1, PRE 1.1, PRE 1, PRE 1.2, PRE 1.4, PRE 1.5, PRE 2, Chinscy Chinczycy}. In the present case, concentrations of $\text{P}$ ($c_{\pi}$), $\text{A}$ ($c_{\alpha}$), and $\text{B}_i$, ($\xi_{i}$), $i \in \mathbb{N}$, are the state variables. Smoluchowski coagulation equations have to be supplemented with the terms related to steps (i), (ii), and (iv). In effect, the time evolution of the system is given by the following rate equations
\begin{eqnarray}
\dot{c}_{\pi} & = & - \tilde{k}_{\pi} c_{\pi},  
\label{set of a b c chemical kinetic equations eq p} 
 \end{eqnarray}
\begin{eqnarray}
\dot{c}_{\alpha} & = & \tilde{k}_{\pi} c_{\pi} - \tilde{k}_{\alpha} c_{\alpha} - \sum_{j=1}^{\infty} \tilde{R}_{j}\xi_{j} c_{\alpha},
\label{complete rate equations alpha}
\end{eqnarray}
\begin{eqnarray}
\dot{\xi}_{1} & = &  \tilde{k}_{\alpha} c_{\alpha} - \tilde{R}_{1}\xi_{1}c_{\alpha}-\sum_{j=1}^{\infty} \left[K_{1j} \xi_{1}\xi_{j} - F_{1j} \xi_{1+j}\right], 
\label{complete rate equations monomers}
\end{eqnarray}
\begin{eqnarray}
\dot{\xi}_{k} & = & \left(\tilde{R}_{k-1} \xi_{k-1}-\tilde{R}_{k} \xi_{k}\right) c_{\alpha} + \sum_{ij} \frac{1}{2} \left[K_{ij} \xi_{i}\xi_{j} -  F_{ij} \xi_{k}\right]  \nonumber \\ &-&  \sum_{j, k+j \leq J} \left[K_{kj} \xi_{k}\xi_{j} - F_{kj} \xi_{k+j}\right],~~~~~~~~~~  k > 1,  
\label{complete rate equations s mers}
\end{eqnarray}
where $\tilde{k}_{\pi} \equiv c_{\rho} k_{\pi}$, $\tilde{k}_{\alpha} \equiv c_{\rho} k_{\alpha}$, and $\tilde{R}_i \equiv c_{\rho} R_i$ are observable reaction rate constants for steps (iv), (i), and (ii), respectively, while $k_{\pi}$, $k_{\alpha}$, and $R_i$ are the corresponding bare rate constants, and $c_{\rho}$ is a concentration of the reducing agent, assumed here to be time-independent. $K_{ij} = K_{ji}$ and $F_{ij} = F_{ji}$ denote coagulation and fragmentation kernels. The first sum in (\ref{complete rate equations s mers}) is restricted to $i+j=k$. 

Initial conditions of interest here are $c_{\pi}(0) \equiv c_0$, $c_{\alpha}(0) \equiv d_0$, with $c_0 + d_0 \equiv q_0 \neq 0$, and  $\xi_{i}(0) = 0$, $i \in \mathbb{N}$. The solution for $c_{\pi}(t)$ is immediate: $c_{\pi}(t) = c_{0}\exp(-\tilde{k}_{\pi} t)$. The original WF model corresponds to $c_{0} = 0$.
 
It follows that $\dot{c}_{\pi}(t) + \dot{c}_{\alpha}(t) + \sum_{j=1}^{\infty} j \dot{\xi}_{j}(t)=0$ (mass conservation constraint). Therefore, the quantity
\begin{equation}
q(t) \equiv c_{\pi}(t) + c_{\alpha}(t) + \sum_{j=1}^{\infty} j \xi_{j}(t) = q_0
\label{the constant of motion}
\end{equation}
is conserved during the time evolution. 

Apart from the assumption of constant concentration of the reducing agent,  $c_{\rho}(t) = c_{\rho}(0)$, ($\mathcal{A}1$), it was also assumed that no source term (no injection mechanism) for either $\text{P}$ or $\text{A}$ species is present ($\mathcal{A}2$); that both the autocatalytic $\text{P}$ to $\text{A}$ reduction reaction, $\text{P} + \text{B}_i  \rightarrow \text{A} + \text{B}_i$ ($\mathcal{A}3$), as well as the disproportionation reaction $\text{P} + \text{B}_1  \longleftrightarrow   2 \text{A}$ ($\mathcal{A}4$) may be neglected. It was also assumed that neither $K_{ij}$, nor $F_{ij}$ depend on concentration of $c_{\rho}$, $c_{\pi}$ or $c_{\alpha}$ ($\mathcal{A}5$), and finally, that chemical species $\text{P}$,   $\text{A}$, and reducing agent ($\text{R}$) do not form clusters ($\mathcal{A}6$). Any of the above assumptions may be abandoned, leading to a generalization of the model defined by Eqs. (\ref{complete rate equations alpha})-(\ref{complete rate equations s mers}) \footnote{An article is in preparation, proposing a model with assumptions ($\mathcal{A}1$)-($\mathcal{A}6$) abandoned.}.

If $\tilde{R}_i = 0$, for all $i\in \mathbb{N}$, (\ref{complete rate equations monomers}) and (\ref{complete rate equations s mers}) become the standard Smoluchowski equations, with the monomer source term $\tilde{k}_{\alpha} c_{\alpha}(t)$ for $\tilde{k}_{\alpha}\neq 0$. However, for $\tilde{R}_i \neq 0$ such reduction is no longer possible, and Eqs. (\ref{set of a b c chemical kinetic equations eq p})-(\ref{complete rate equations s mers}) with their generalizations are members of a wider class of 'reaction-aggregation' equations. Few models of this type have been found in the literature \cite{PRE 1.4, PRE 1.5, PRE 2}.

\textit{Method of moments.} To analyze properties of Eqs. (\ref{complete rate equations alpha})-(\ref{complete rate equations s mers}), it is convenient to apply the standard method of moments. The $\mu$-th moment of the cluster mass distribution is defined as $M_{\mu}(t) = \sum_{j=1}^{\infty} j^{\mu}  \xi_{j}(t)$. From Eqs. (\ref{complete rate equations monomers}) and (\ref{complete rate equations s mers}) we obtain
\begin{equation}
\dot{M}_{\mu} = \tilde{k}_{\alpha} c_{\alpha} + \sum_{j=1} \mathcal{G}^{(\mu)}_{j} \xi_{j} c_{\alpha} + \sum_{p, q}\mathcal{S}^{(\mu)}_{p q} \xi_{p} \xi_{q} + \sum_{p=2} \mathcal{T}^{(\mu)}_{p} \xi_{p},
\label{moments general explicite M mu}
\end{equation}
 where $\mathcal{S}^{(\mu)}_{p q} = \mathcal{S}^{(\mu)}_{q p} \equiv \frac{1}{2}\big((p+q)^{\mu} - p^{\mu} - q^{\mu}\big) K_{pq}$, $\mathcal{T}^{(\mu)}_{p} \equiv \sum_{i=1}^{p-1} \big(i^{\mu} - \frac{1}{2}p^{\mu}\big) F_{i,p-i}$ and  $\mathcal{G}^{(\mu)}_{j} \equiv [(j+1)^{\mu}-j^{\mu}] \tilde{R}_{j}$.  

From Eq. (\ref{moments general explicite M mu}) we see, 
first, that all moments grow due to the monomer production (i), note, $\tilde{k}_{\alpha} c_{\alpha} \geq 0$. Secondly, for $\mu=0$, $\mathcal{G}^{(0)}_{j}=0$, because the total cluster concentration $M_0$ is not affected by autocatalytic reaction (ii). For $\mu=1$, $\mathcal{S}^{(1)}_{p q}=\mathcal{T}^{(1)}_{p}=0$, i.e., the total cluster mass is not changed by coagulation or fragmentation. Thus, for $\mu=1$, stationary solution of (\ref{moments general explicite M mu}) exists for $\tilde{k}_{\alpha}>0$. Indeed, for $(\tilde{k}_{\alpha} + \sum_{j=1} \mathcal{G}^{(1)}_{j} \xi_{j}) > 0$, from $\dot{M}_1 = 0$ follows $c_{\alpha} = 0$. Since $\lim_{t \to \infty} c_{\pi}(t) =0$, Eq. (\ref{the constant of motion}) gives $\lim_{t \to \infty} M_{1}(t) \equiv \bar{M}_{1} = q_0$, as expected for an irreversible reaction. Additionally, for $K_{ij} = F_{ij}=0$, when $\mathcal{S}^{(\mu)}_{p q}=\mathcal{T}^{(\mu)}_{p}=0$, similar procedure proves that all $M_{\mu}$ approach stationary value.

In order to obtain tractable system of time-evolution equations for $M_{\mu}(t)$ and $c_{\alpha}$, a restriction is imposed on $\mu$ and $F_{ij}$ so that $\mu \in \mathbb{N}\cup 0$, $F_{ij}\equiv 0$, and $\tilde{R}_{i}$ and $K_{ij}$ given by
\begin{equation}
K_{ij}= \kappa_{0} + \kappa_{1}   (i+j) + \kappa_{2}  ij, ~~~~\tilde{R}_{i} = \tilde{a}_{R} i + \tilde{b}_{R},
\label{R od i}
\end{equation}
where $\tilde{a}_{R}$, $\tilde{b}_{R}$, $\kappa_{0}$, $\kappa_{1}$, $\kappa_{2}$ are arbitrary non-negative coefficients. Under these simplifying assumptions, Eqs. (\ref{complete rate equations alpha}) and (\ref{moments general explicite M mu}) for $\mu=0, 1, 2$ assume the form 
\begin{eqnarray}
\dot{c}_{\alpha} & = & \tilde{k}_{\pi} c_{\pi}(t) - \tilde{k}_{\alpha} c_{\alpha} - \tilde{a}_{R} M_1 c_{\alpha} - \tilde{b}_{R} M_0 c_{\alpha}, 
\label{moments general restricted K R alpha} \\
\dot{M}_0 & = & \tilde{k}_{\alpha} c_{\alpha} - \frac{1}{2} \kappa_{0} M^{2}_{0} - \kappa_{1} M_0 M_1  -  \frac{1}{2} \kappa_{2} M^{2}_{1}, 
\label{moments general restricted K R M0} \\
\dot{M}_1 & = & \tilde{k}_{\alpha} c_{\alpha} +  \tilde{a}_{R} M_1 c_{\alpha} + \tilde{b}_{R} M_0 c_{\alpha},
\label{moments general restricted K R M1} \\
\dot{M}_2 & = & \tilde{k}_{\alpha} c_{\alpha} +  \big[2 \tilde{a}_{R} M_2 + (\tilde{a}_{R} + 2 \tilde{b}_{R}) M_1 + \tilde{b}_{R} M_0\big]c_{\alpha} \nonumber \\ &+& \kappa_{0} M^{2}_{1} + 2 \kappa_{1} M_1 M_2  + \kappa_{2} M^{2}_{2}.
\label{moments general restricted K R M2} 
\end{eqnarray}
The initial conditions, $\forall k: \xi_k(0)=0$ give $M_{\mu}(0) = 0$, $\forall \mu$,  while variables $c_{\alpha}$ and $M_1$ are not independent, since they obey the constraint (\ref{the constant of motion}) for  $M_1 = \sum_{j=1}^{\infty} j \xi_{j}(t)$.

Explicit form of the corresponding evolution equations for higher moments ($\mu > 2$) may be easily found using Eq. (\ref{moments general explicite M mu}). However, higher moments are not needed to calculate two basic characteristics of the cluster mass distribution, i.e., mean cluster size, $\langle i \rangle$, and variance, $\sigma^2(i)$, as given by 
\begin{equation}
\langle i \rangle  = \frac{M_1}{M_0},  ~~~~~~\sigma^2(i) = \frac{M_2}{M_0} - \left(\frac{M_1}{M_0}\right)^2.
\label{mean size}
\end{equation}
\textit{Absence of coagulation.} The case of negligible coagulation and fragmentation is at the center of our interest. In colloidal systems, experimentally, this is achieved by addition of a stabilizing agent, e.g. the polyvinyl alcohol (PVA) or polyvinylopyrrolidone (PVP), which inhibits coagulation, affecting chemical reactions to a lesser extent \cite{Our group glucose, Our group ascorbic acid}. Models of autocatalytic reaction without coagulation may also provide an adequate description for other systems, e.g. simple chain polymers. 

For $K_{ij} = F_{ij} = 0$,  Eq. (\ref{complete rate equations s mers}) divided by Eq. (\ref{complete rate equations monomers}) gives
\begin{eqnarray}
\frac{d\xi_{k}}{d\xi_{1}} & = & \frac{\tilde{R}_{k-1} \xi_{k-1} - \tilde{R}_{k} \xi_{k}}{\tilde{k}_{\alpha} -  \tilde{R}_{1} \xi_{1}}. 
\label{s moments general explicite ksi k particular divided by ksi 1 no coag}
\end{eqnarray}
Eqs. (\ref{s moments general explicite ksi k particular divided by ksi 1 no coag}) for $k=2, 3 \ldots$ form a system of coupled linear differential equations, allowing to determine each $\xi_{k}$ as a function of $\xi_{1}$, $\xi_{k} = s_k(\xi_{1})$. If coagulation and fragmentation processes could be neglected, Eqs. (\ref{s moments general explicite ksi k particular divided by ksi 1 no coag}), therefore all $s_k(\xi_{1})$ functions, for given initial conditions, depend only on $\tilde{k}_{\alpha}$ and $\tilde{R}_i$ parameters, and not on chemical reactions and physical process, e.g. injection of $\text{P}$ and $\text{A}$ substrates, unless they involve $B_i$ clusters. In particular, $s_k(\xi_{1})$, hence the stationary values of $\xi_{k}$,  $\bar{\xi}_{k} \equiv  \lim_{t \to \infty}\xi_{k}(t)$ do not depend on assumptions $\mathcal{A}1$-$\mathcal{A}3$.

The above holds true for any equation derived by dividing  Eq. (\ref{complete rate equations monomers}), or one of Eqs. (\ref{moments general restricted K R alpha})-(\ref{moments general restricted K R M2}), by Eq. (\ref{moments general restricted K R M0}) or (\ref{moments general restricted K R M1}), which allow expressing $\xi_1$ and $M_{\mu}$ as the functions of $M_{0}$ or $M_{1}$ only, see Eqs. (\ref{moments linear R division M1 M0}), (\ref{moments constant R division M1 M0}), and (\ref{constant R division ksi 1 M0}) below.

As a consequence, all special cases or generalizations of the model defined by Eqs. (\ref{set of a b c chemical kinetic equations eq p})-(\ref{complete rate equations s mers}) with the monomer production provided solely by (i), and with identical mechanism of autocatalytic reaction (ii) belong to the same universality class.

Two special cases of the reaction kernel (\ref{R od i}) are analyzed below: the linear kernel,  proportional to the cluster mass, $\tilde{R}_{i} \propto i$, $\tilde{b}_{R} = 0$; and the size-independent one, $\tilde{a}_{R} = 0$.

\textit{Linear reaction kernel.} For $\kappa_0 = \kappa_1 = \kappa_2 = 0$, $\tilde{b}_{R} = 0$, and $\tilde{a}_{R} \neq 0$, Eqs. (\ref{set of a b c chemical kinetic equations eq p}), (\ref{moments general restricted K R alpha}), and (\ref{moments general restricted K R M1}) become identical to the rate equations analyzed in Refs. \cite{Our group glucose, BWKEIK} ($c_0 \neq 0$) and \cite{WF 1, WF 2,  WF 3, WF protein aggregation} ($c_0 = 0$). These equations have the same form for arbitrary choice of $K_{ij}$ and $F_{ij}$, hence the reaction rate of (ii), proportional to the total mass of the clusters ($M_1$), does not depend on presence of the coagulation or fragmentation. 

$c_{\alpha}$ may be eliminated in favor of $c_{\pi}(t)$ and $M_1 \equiv x$ using Eq. (\ref{the constant of motion}), which yields the form of Eqs. (\ref{moments general restricted K R M1}) 
\begin{equation}
\dot{x}= (\tilde{k}_{\alpha} + \tilde{a}_{R} x)\left(f(t) - x\right),
\label{pre-Bernoulli o d e for x}
\end{equation}
where $f(t) \equiv q_0 - c_{\pi}(t) = d_0 + c_{0}(1-\exp(-\tilde{k}_{\pi} t))$ and $0 \leq x \leq q_0$. The following substitution: $u = \tilde{k}_{\alpha} + \tilde{a}_{R} x$ transforms (\ref{pre-Bernoulli o d e for x}) into Bernoulli-type equation, which gives 
\begin{equation}
x(t) =  \frac{1}{\tilde{a}_{R}} e^{\Phi(t)} \left(  \frac{1}{\tilde{k}_{\alpha}}  + \int_{0}^{t} e^{\Phi(\eta)} d \eta \right)^{-1}  - \frac{\tilde{k}_{\alpha}}{\tilde{a}_{R}}, 
\label{x of t first encounter integral form}
\end{equation}
where $\Phi(t) \equiv C_1 (\exp(-\tilde{k}_{\pi} t) - 1) + C_2 t$, $C_2 = \tilde{a}_{R} q_0 + \tilde{k}_{\alpha}$, and $C_1 = c_0\tilde{a}_{R}/\tilde{k}_{\pi} = c_0 k_{4}/k_{1}$. For the original WF model, where $c_{0} = 0$, $C_1 = 0$, and $f(t) = d_{0} \neq 0$, (\ref{x of t first encounter integral form}) reduces to  
\begin{equation}
x_{\alpha \beta}(t) =  -\left(d_0 + \frac{\tilde{k}_{\alpha}}{\tilde{a}_{R}}\right)\cdot \left(\frac{\tilde{k}_{\alpha}}{d_0 \tilde{a}_{R}} e^{(\tilde{a}_{R}d_0 + \tilde{k}_{\alpha})t} + 1  \right)^{-1}  + d_0,
\label{x of t for 2 step WF}
\end{equation}
where $ x_{\alpha \beta}(0) = 0$ and $\lim_{t \to \infty} x_{\alpha \beta}(t) = d_0$  \cite{WF 1, BWKEIK}. However, for $c_{0} \neq 0$, $x(t)$ cannot be expressed as a combination of finite number of elementary functions. Still Eq. (\ref{x of t first encounter integral form}) may be given a more convenient form $x(t) = \chi(z(t))$, where $z(t) \equiv C_1 \exp(-\tilde{k}_{\pi} t)$  and $\chi(z)$ is given by 
\begin{eqnarray}
\chi(z) &=& \frac{e^{z}}{ \tilde{a}_{R}}\Bigg(\Bigg[\frac{e^{C_1}}{\tilde{k}_{\alpha}} + \frac{{}_{1}F_1(1 - \gamma;2 - \gamma;C_1)}{\tilde{k}_{\pi}(1 - \gamma)} \Bigg] \frac{z^{\gamma-1}}{C_1^{\gamma-1}} \nonumber \\ &-& \frac{{}_{1}F_1(1 - \gamma;2 - \gamma;z)}{\tilde{k}_{\pi}(1 - \gamma)} \Bigg)^{-1}   - \frac{\tilde{k}_{\alpha}}{\tilde{a}_{R}}. 
\label{x tilde of z}
\end{eqnarray}
Above, $\gamma = 1 + C_2/\tilde{k}_{\pi}$ and ${}_{1}F_1(a;b;z)$ denote confluent hypergeometric function \cite{AS}. 

Since in this case, there is an analytic solution (\ref{x tilde of z}) for $M_1(t) = x(t)$, it is convenient to express both $M_0$ and $M_2$ as the functions of $M_1$. 

Eq. (\ref{moments general restricted K R M0}) divided by Eq. (\ref{moments general restricted K R M1}) gives 
\begin{eqnarray}
\frac{d M_0}{d M_1} & = & \frac{\tilde{k}_{\alpha}}{\tilde{k}_{\alpha} + \tilde{a}_{R} M_1} =\frac{1}{1 + \frac{\omega}{q_0} M_1},
\label{moments linear R division M1 M0}
\end{eqnarray}
where $\omega = q_0\tilde{a}_{R}/\tilde{k}_{\alpha}=q_0 a_{R}/k_{\alpha}$. Eq. (\ref{moments linear R division M1 M0}) gives
\begin{eqnarray}
M_0(M_1) & = & \frac{q_0}{\omega} \ln\left( 1 + \frac{\omega}{q_0} M_1 \right). 
\label{moments linear R M1 M0 as a function}
\end{eqnarray}
In parallel, dividing Eq. (\ref{moments general restricted K R M2}) by Eq. (\ref{moments general restricted K R M1}) gives  
\begin{eqnarray}
M_2(M_1) & = & M_1 \left(1 + \frac{\omega}{q_0} M_1 \right). 
\label{moments linear R M1 M2 as a function}
\end{eqnarray}
Using Eqs. (\ref{moments linear R M1 M0 as a function}) and (\ref{moments linear R M1 M2 as a function}), explicit formulas for mean cluster size and variance (\ref{mean size}) may be easily derived.
 
Solving Eqs. (\ref{s moments general explicite ksi k particular divided by ksi 1 no coag}) for $\tilde{R}_i =\tilde{a}_{R} i$ gives 
\begin{equation}
\xi_{k}(t)= s^{(a)}\big(\xi_{1}(t)\big) = \frac{1}{k} \frac{q_0}{\omega}  \left(\frac{\omega}{q_0} \xi_{1}(t) \right)^{k}, ~~~~k\geq 1.
\label{ksi k od ksi 1 linear}
\end{equation}
In order to find $\xi_{1}(t)$, $M_0$ (\ref{moments linear R M1 M0 as a function}) is equated with  $\sum_{j=1}^{\infty} \xi_{j}(t)$. Employing  (\ref{ksi k od ksi 1 linear}) and the identity
\begin{equation}
\sum_{k=1}^{\infty} \frac{x^k}{k} = \ln\left(\frac{1}{1-x}\right) = \ln\left(1+\frac{x}{1-x}\right),
\end{equation}
where $x=\Omega \cdot (\Omega + 1)^{-1}$ and $\Omega = \omega M_1/q_0$, gives 
\begin{equation}
\xi_1(t) = h^{(a)}_1\big(M_1(t)\big) = \frac{M_1(t)}{1+\frac{\omega}{q_0}M_1(t)}.
\label{ksi 1 od M 1 linear R}
\end{equation} 
Combining Eqs. (\ref{ksi k od ksi 1 linear}) and (\ref{ksi 1 od M 1 linear R}), the following is found 
\begin{equation}
\bar{\xi}_k  \equiv \lim_{t \to \infty} \xi_k(t) = \frac{1}{k} \frac{q_0}{\omega}  \left(\frac{\omega}{\omega + 1}\right)^k.
\label{ksi k linear R stationary}
\end{equation} 
Finally, for $c_0 = 0$, Eqs. (\ref{x of t for 2 step WF}), (\ref{ksi k od ksi 1 linear}) and (\ref{ksi 1 od M 1 linear R}) yield  
\begin{equation}
\xi^{(\alpha \beta)}_{k}(t) = \frac{1}{k} \frac{d_0}{\omega} \left(\frac{\omega}{\omega + 1}\right)^k \left(1 - e^{-\tilde{k}_{\alpha}(1+ \omega)t}\right)^{k}.
\label{ksi 1 of t for 2 step WF}
\end{equation}

\textit{Size-independent reaction kernel.} For colloidal systems, $\tilde{R}_{j}$ given by Eq. (\ref{R od i}), for $\tilde{a}_{R} = 0$ and $\tilde{b}_{R} \neq 0$, provides a lower bound for any realistic functional form of $\tilde{R}_{j}$, which is expected to be a non-decreasing function of $j$. Moreover, in the absence of coagulation, i.e., for $\kappa_0 = \kappa_1 = \kappa_2 = 0$,  Eqs. (\ref{moments general restricted K R alpha})-(\ref{moments general restricted K R M2}) may be regarded as a simple model of linear polymer growth \footnote{In this case, $\text{A}$ denotes an 'active monomer', while $ \text{B}_1$ is an 'inert' or 'inactive monomer' which cannot take part in the polymerization process.}, cf. \cite{PRE 2}. In this case, to solve Eqs. (\ref{moments general restricted K R alpha})-(\ref{moments general restricted K R M2}), Eq. (\ref{moments general restricted K R M1}) is divided by Eq. (\ref{moments general restricted K R M0}), giving
\begin{eqnarray}
\frac{d M_1}{d M_0} & = & 1 + \frac{\tilde{b}_{R}}{\tilde{k}_{\alpha}} M_0 = 1 + \frac{\omega}{q_{0}} M_0,
\label{moments constant R division M1 M0}
\end{eqnarray}
where $\omega \equiv q_0 \tilde{b}_{R}/\tilde{k}_{\alpha} = q_0 b_{R}/k_{\alpha}$. Eq. (\ref{moments constant R division M1 M0}) gives
\begin{eqnarray}
M_1(M_0) & = & \frac{\omega}{2 q_0} M^2_0 + M_0. 
\label{moments constant R M1 M0 as a function}
\end{eqnarray}
Applying a parallel procedure to Eqs. (\ref{moments general restricted K R M2}) and (\ref{moments general restricted K R M0}), with the use of $M_1(M_0)$ (\ref{moments constant R M1 M0 as a function}), and solving the equation gives
\begin{eqnarray}
M_2(M_0) & = & M_0 +  \frac{3\omega}{2q_{0}} M^2_0 + \frac{\omega^2}{3 q_{0}^{2}} M^3_0. 
\label{moments constant R M2 M0 as a function}
\end{eqnarray}
Using Eqs. (\ref{moments constant R M1 M0 as a function}), (\ref{moments general restricted K R M1}) and the constraint (\ref{the constant of motion}) in order to eliminate $c_{\alpha}$, yields the following equation
\begin{eqnarray} 
\dot{M}_0 & = & \tilde{k}_{\alpha} \left(q_0 - c_{\pi}(t)  - M_0 - \frac{\omega}{2 q_0} M^2_0 \right).  
\label{moments constant R no coagulation M0 solo}
\end{eqnarray}

For $c_0 \neq 0$, it was impossible to find an analytical solution of Eq. (\ref{moments constant R no coagulation M0 solo}). However, the stationary solution of Eq. (\ref{moments constant R no coagulation M0 solo}) may easily be found. Combining Eq. (\ref{moments constant R M1 M0 as a function}) and the limit $\lim_{t \to \infty} M_1 = q_0$, gives 
\begin{equation}
\lim_{t \to \infty} \left[M_0(t) + \frac{\omega}{2 q_0} M^2_0(t)\right] = \bar{M}_{0} + \frac{\omega}{2 q_0}\bar{M}^2_{0} = q_0.
\label{moments constant R no coagulation M0 quadratic}
\end{equation} 
Eq. (\ref{moments constant R no coagulation M0 quadratic}) implies
\begin{equation}
\bar{M}_{0} = \frac{q_0}{\omega} \left( \sqrt{1 + 2 \omega}    - 1 \right).
\label{moments constant R no coagulation M0 stationary}
\end{equation}

Eq. (\ref{moments constant R no coagulation M0 solo}) may also be easily solved for $c_0 = 0$, giving
%
\begin{eqnarray} 
M_0(t) & = & \frac{d_{0}}{\omega} \cdot \frac{ \left(\eta-1\right)\left(1- e^{- \tilde{k}_{\alpha} \eta t}\right) }{1 + \left(\frac{\eta-1}{\eta+1}\right) e^{- \tilde{k}_{\alpha} \eta t}}, 
\label{moments constant R no coagulation M0 solo WF two step solution}
\end{eqnarray}
%
where $\eta \equiv \sqrt{1 + 2 \omega}$. 
$M_0(t)$ (\ref{moments constant R no coagulation M0 solo WF two step solution}) has the following properties: $ M_0(0) = 0$, $ M_0(t) < \bar{M}_{0}$, where $\bar{M}_{0}$ is given by (\ref{moments constant R no coagulation M0 stationary}) and $\lim_{t \to \infty} M_0(t) = \bar{M}_{0}$, as might have been expected. Eqs. (\ref{moments constant R M1 M0 as a function}), (\ref{moments constant R M2 M0 as a function}), and (\ref{moments constant R no coagulation M0 solo WF two step solution}) may be used to calculate the time-dependence of $\langle i \rangle_p$ and $\sigma^2(i)$ (\ref{mean size}).

For $\tilde{R}_i =\tilde{b}_{R}$, solutions of Eqs. (\ref{s moments general explicite ksi k particular divided by ksi 1 no coag}) read 
\begin{equation}
\xi_{k}=s^{(b)}(\xi_{1}) = \frac{q_0}{\omega}\left\{ 1 + u(\xi_{1}) \text{W}_{k-1}\left[\ln\left(\frac{-1}{u(\xi_{1})}\right)\right] \right\}, 
\label{ksi k od ksi 1 constant}
\end{equation}
where $u(\xi_{1}) = \frac{\omega}{q_0} \xi_{1} - 1$ and  $W_{i}(x) \equiv \sum_{j=0}^i x^j/j!$, which may be easily verified (note, $W_{k-1}(x)=W^{\prime}_{k}(x)$).
To obtain explicit form of $\xi_1(t)$, Eq. (\ref{s moments general explicite ksi k particular divided by ksi 1 no coag}) is divided by Eq. (\ref{moments general restricted K R M0}). This yields 
\begin{eqnarray}
\frac{\dot{\xi_{1}}}{\dot{M_0}} & = & \frac{d \xi_{1}}{d M_0} =  1 - \frac{\omega}{q_0} \xi_{1}.
\label{constant R division ksi 1 M0}
\end{eqnarray}
The solution of Eq. (\ref{constant R division ksi 1 M0}) is given by
\begin{equation}
\xi_1(t) = h^{(b)}_0\big(M_0(t)\big) = \frac{q_0}{\omega} \left( 1- e^{-\frac{\omega}{q_0}  M_0(t)}\right), 
\label{ksi 1 od M 0 constant R}
\end{equation} 
where for $c_0 = 0$, $M_0(t)$ is given by (\ref{moments constant R no coagulation M0 solo WF two step solution}).

Using Eqs. (\ref{moments constant R no coagulation M0 stationary}), (\ref{ksi k od ksi 1 constant}), and (\ref{ksi 1 od M 0 constant R}), stationary value of each $\xi_k$ may easily be found
\begin{equation}
\bar{\xi}_k  \equiv \lim_{t \to \infty} \xi_k = \frac{q_0}{\omega} \left[1- e^{-(\eta - 1)}W_{k-1}(\eta - 1) \right].
\label{ksi 1 od M 0 constant R stationary}
\end{equation} 

Finally, for the cluster size-independent reaction kernel, coagulation process, when present, decrease the rate of the autocatalytic reaction (ii). This becomes intuitively clear, when the present model is applied to describe growth of linear polymers with two active reaction sites at the ends of the polymer chain, since each coagulation event reduces the number of reaction sites by a factor two. Clearly, the influence of fragmentation processes is exactly opposite.

\textit{Summary and discussion.} In this paper, time-evolution rate equations for the model of transition-metal nanocluster formation, as proposed by Watzky and Finke, have been introduced. The equations introduced constitute a natural generalization of both the Smoluchowski coagulation equations, and the rate equations, describing the kinetics of monomer production, autocatalytic nanoparticle surface growth, and other chemical reactions present in the system.

In the absence of coagulation and fragmentation, exact solutions of the model equations have been found for the autocatalytic rate constant (reaction kernel) proportional to cluster mass, $\tilde{R}_i \propto i$, as well as for the cluster-size independent one, $\tilde{R}_i = \text{const}$.

Secondly, it was demonstrated that the functional dependence of the $k$-atom cluster concentrations, $\xi_{k}$ on $\xi_{1}$, given by $\xi_{k}=s(\xi_{1})$, and  $\bar{\xi}_k \equiv \lim_{t \to \infty} \xi_{k}(t)$, are completely determined by the assumed model mechanism of monomer production and autocatalytic growth, but do not depend on any other chemical reaction. However, this does not hold for the  $\xi_{k}(t)$ functions.

In conclusion,  kinetically inequivalent generalizations of the present model are divided into universality classes, whereby two such models belong to the same class when they yield the same $s(\xi_{1})$ function.
 
\textit{Acknowledgements.} I would like to thank Krzysztof Fitzner,  Pawe\l~ G\'{o}ra, Wiktor Jaworski, Piotr Mierzwa, Krzysztof Pac\l awski, and Bartek Streszewski for inspiring discussions and acknowledge their beneficial influence during my work on this paper.

{}

\end{document}